\documentclass{jpsj2}
\title{Long-term power-law fluctuation in Internet traffic}
\author{Shin-ichi \textsc{Tadaki}}
\inst{Computer and Network Center, Saga University, Saga 840-8502}

\abst{%
Power-law fluctuation in observed Internet packet flow are discussed.
The data is obtained by a multi router traffic grapher (MRTG) system
for 9 months.  The internet packet flow is analyzed using the detrended
fluctuation analysis.  By extracting the average daily trend, the data
shows clear power-law fluctuations.  The exponents of the fluctuation
for the incoming and outgoing flow are almost unity.  Internet
traffic can be understood as a daily periodic flow with power-law
fluctuations.}
\kword{Internet traffic, DFA, MRTG, power-law fluctuation}

\begin{document}
\maketitle

\section{Introduction}\label{Introduction}
%Research on Internet Traffic
The Internet is one of the most important modern infrastructures for
daily communication.  On the contrary to the importance of the
Internet, it has no global centers for controlling the global
structure and the data transmission. It is namely an autonomous
growing network.  Complex properties in various autonomous growing
networks have been studied in the viewpoints of small-world and
scale-free properties.  The structural and transport properties of the
Internet and services such as the world wide web (WWW) also have been
attracting scientific interests from the viewpoints of statistical
physics\cite{PastorSatoras,KG2005}.

%Overview of researches on time sequences
Internet traffic had been thought to be modeled by a Poisson process,
because hosts are assumed to send data packets randomly.  The validity
of this assumption has clearly lost on the basis of various
experimental measurements\cite{Paxson}.  Power-law properties of
Internet traffic have been investigated instead.  Csabai investigated
a time series of round trip time (RTT) for two weeks and observed its
self-similarity\cite{Csabai}.  Takayasu, Takayasu and Sato observed a
time series of RTT and packet density fluctuations of three days long
for discussing power-law properties of Internet packet
flow\cite{Takayasu:1994}.  In smaller time scales than one second, on
the other hand, Internet traffic was reported to be almost white
noise\cite{Ribeiro}.

The Internet is a man-made communication system.  So the traffic on
the Internet is affected by human social activities.  It may contain daily
periodicity corresponding to periodicities of activities in human
societies.  To avoid effects of social periodic activities, researches
on Internet traffic mentioned above focus their attentions to smaller
time-scales than a day.  The existence of power-law correlations
longer than several days is also interesting in itself.

Components of the Internet, for example, Ethernets, routers and the
transport control protocol (TCP), have exclusion and queuing
mechanisms.  The Ethernet protocol has the carrier sense multiple
access/collision detective (CSMA/CD) mechanism\cite{IEEE802}, which
requires the binary exponential back-off (queuing) algorithm for hosts
connecting to an Ethernet.  A router has a finite size of queue for
forwarding packets using the Internet protocol (IP). TCP contains a
congestion control mechanism to change the flow rate.  TCP also
requires a sender to re-send packets for fail-safe data transmission.
These queuing effects are pointed to be a key feature of power-law
behavior observed in the Internet\cite{Takayasu:1998,Fukuda:2000}.

%Relation to vehicle traffic

We can also observe power-law fluctuations in vehicle traffic.
Vehicle traffic in an expressway also has exclusion and queuing
mechanisms.  A traffic lane and the finiteness of car length have
exclusion effects on traffic flow.  The density fluctuation in traffic
flow, such as traffic jam, propagates upstream by the exclusion
effect.  The simplest model of traffic flow is a cellular automaton
(CA) model, such as Wolfram's rule 184 CA\cite{Wolfram}, which
includes the exclusion effect.  Some models of traffic flow reproduce
power-law behavior\cite{Takayasu1993,Yukawa1996,Tadaki1999}.  In
observations of real expressway traffic, Mush and Higuch observed
$1/f$ fluctuations within smaller time-scale than several
hours\cite{MH76}.

Expressway traffic is non-stationary flow consisting of correlations
with various time scales.  Traffic flow also contains daily
periodicity reflecting human social activities.  One of methods for
analyzing non-stationary time series is the detrended fluctuation
analysis (DFA)\cite{PBHSSG94,PHSG95}.  By extracting the average daily
periodicity, power-law fluctuations in expressway traffic are found
to be extended longer than several months using DFA\cite{Tadaki2006}.

The purpose of this paper is to investigate large time-scale behavior
of Internet traffic. We analyze Internet packet flow passing through a
network gateway observed using a multi router traffic grapher (MRTG)
system\cite{MRTG}.  Applying DFA method on the data, the long-range
correlation will be discussed.

The organization of this paper is as follows.  DFA is described
briefly in \S~\ref{DFA}.  We analyze Internet packet flow observed
using MRTG.  Section \ref{TheData} describes the observed data and the
result using DFA.  As in our previous work on vehicle traffic, a
modified data is defined by extracting the daily average flow from the
raw data.  The modified data is analyzed in \S~\ref{ModData}.  Section
\ref{Summary} is devoted to summary and discussion.

\section{Detrended Fluctuation Analysis}\label{DFA}
The detrended fluctuation analysis (DFA) is one of methods for
analyzing non-stationary time series.  It was first developed for
analyzing the long-range correlation in deoxyribonucleic acid (DNA)
sequences\cite{PBHSSG94,PHSG95}.  The method has been applied to
various time series with non-stationarity for analyzing their
power-law properties.  The theoretical properties of the method have
also been discussed\cite{KKRHB01,VGMHB01}.

The simplest form of the method is described as follows. Consider to
analyze a raw temporal data $\lbrace u(t)\rbrace$ ($0\le t<T$).
First, the \textit{profile} $y(t)$ of the raw temporal data $\lbrace
u(t)\rbrace$ ($0\le t<T$) is defined as the accumulated deviation from
the average
\begin{equation}
y(t)=\sum_{i=0}^{t}\left[u(i)-\langle u\rangle\right],
\end{equation}
where $\langle u\rangle=T^{-1}\sum_{t=0}^{T-1}u(t)$ is the temporal
average value of the raw data $\lbrace u(t)\rbrace$.

The entire time sequence of the profile $y(t)$ of length $T$ is
divided into $T/l$ non-overlapping segments of length $l$.  The
\textit{local trend} $\tilde{y}_n(t)$ in the $n$-th segment is 
defined by fitting the raw profile $y(t)$ in the segment.  We here
employ the linear least-squares method to fit the profile.  This is
called \textit{first-order} DFA.

The \textit{detrended profile} $y_l(t)$ is defined as the deviation of
the original profile $y(t)$ from the local trend $\tilde{y}_n(t)$
\begin{equation} 
y_l(t) =y(t) - \tilde{y}_n(t),\quad \mathrm{if\ }nl\le t<(n+1)l.  
\end{equation} %

The standard deviation of the detrended sequence is defined as the
mean square of the detrended profile
\begin{equation}
F^2(l)=\frac{1}{T}\sum_{t=0}^{T-1}y_l^2(t).\label{Deviation:1}
\end{equation}
By analyzing the dependence of the standard deviation $F(l)$ on the
segment length $l$, we find the long-range correlation in the
non-stationary time sequences.  If the standard deviation $F(l)$
behaves as a power of the segment length $l$
\begin{equation}
F(l)\sim l^{\alpha},
\end{equation}
the power spectrum $P(k)$ of the time sequence $u(t)$ also obeys
the power-law
\begin{equation}
P(k)\sim k^{-\beta},\quad \beta=2\alpha-1.
\end{equation}

\section{Analysis of data}\label{TheData}

 \begin{figure}[tb]
\begin{center}
\resizebox{0.4\textwidth}{!}{\includegraphics{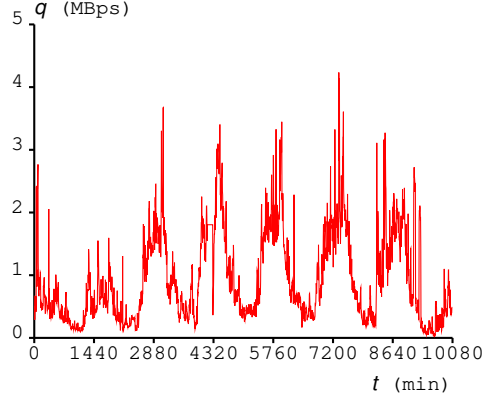}}
\caption{A sample packet flow for one week observed using MRTG.
Data are 5 minute average.  The horizontal axis denotes time (minute)
and the vertical one does averaged packet flow (Mega Byte per second).
We recognize daily periodicity (1440 min).
}
\label{flow.eps}
\end{center}
\end{figure}

We observe Internet packet flow at the gateway of Saga University to
Kyushu University, where we connect our university local area network
(LAN) to the science information network (SINET), the Japanese
backbone network for academic organizations\cite{SINET}.  The
bandwidth of the line is 100 Mbps. Data of packet flow is obtained as
a five-minute average value using MRTG.  The data set consists of MRTG
packet data observed from September, 2005 to May, 2006.

The period where the data are collected, includes the end of the
second semester in 2005 school year.  And some parts of data are lost
because of troubles in the network system and the MRTG system.  Namely
the data has various types of non-stationarity and defects.

\begin{figure}[tb]
\begin{center}
\resizebox{0.4\textwidth}{!}{\includegraphics{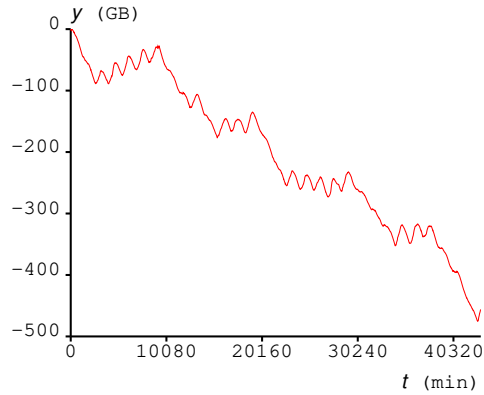}}
\caption{A sample profile of the packet flow corresponding to 
Fig.~\ref{flow.eps}.  The horizontal axis denotes time (minute) and
the vertical one does accumulated packet fluctuation around the
average (Giga Byte).  Steep decreases correspond to inactivity in weekends.}
\label{profile.eps}
\end{center}
\end{figure}

Figure \ref{flow.eps} shows a sample data of the packet flow $q(t)$
for one week observed using MRTG.  Daily periodic behavior (1440
minutes) can be found clearly corresponding to activities in our
university.  The data also contains weekly periodic behavior, which
can be found in the profile (Fig.~\ref{profile.eps}).

\begin{figure}[tb]
\begin{center}
\resizebox{0.4\textwidth}{!}{\includegraphics{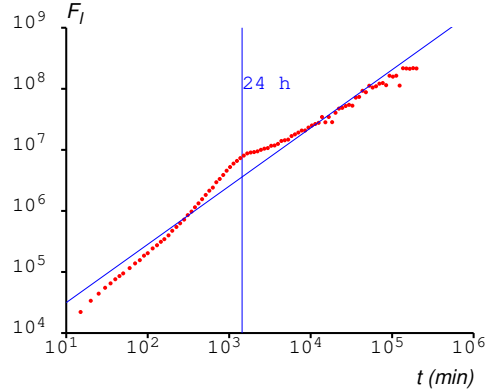}}
\caption{Dependence of the standard deviation $F(l)$ on segment
length $l$ for MRTG data. There is a bending point at one day long.
This is a typical feature of periodic time sequences with power-law
fluctuations.}
\label{dfa.eps}
\end{center}
\end{figure}

We apply the DFA method on the time sequence of the packet flow
$q(t)$.  First the profile $y(t)$ is defined by
\begin{equation}
y(t)=\sum_{i=0}^{t}\left[q(i)-\langle q\rangle\right].
\label{ProfileDefine}
\end{equation}
The profile represents the accumulated packet fluctuation around the
average.  The profile shown in Fig.~\ref{profile.eps} is 30 days
long. We can recognize weekly periodicity (10080 minutes).  The
inactivity of network use in weekends appears as steep decrease of
the curve.

By applying the DFA method, the dependence of the standard deviation
$F(l)$ on the segment length $l$ of local trends is obtained as shown
in Fig.~\ref{dfa.eps}.  The result seems to be divided into two
regions of different exponents.  Two regions are connected at a
bending point of the curve at one day long.

The feature appeared in Fig.~\ref{dfa.eps} is similar to that observed
in the DFA analysis of expressway vehicle traffic\cite{Tadaki2006}.
The curve of the standard deviation $F(l)$ has a bending point, which
corresponds to the daily periodicity of the traffic.  This bending
appeared in the curve of $F(l)$ is a typical feature of periodic time
series with power-law fluctuations\cite{Hu}.  The bending point
corresponds to the dominant periodicity in the time series.  Namely
the Internet traffic observed at the gateway seems to be a daily
periodic time sequence with power-law fluctuations.

The DFA analysis of Internet traffic has been reported by Fukuda,
Nunes Amaral and Stanley\cite{FAS2003}.  They have observed a shorter
time sequence than a day to avoid effects of daily periodicity
reflecting network user's activities.  They have found power-law
fluctuations in the Internet traffic.  Our research extends the DFA
analysis of Internet traffic for longer period than one month.

\section{Modified Data and Analysis}\label{ModData}
\begin{figure}[tb]
\begin{center}
\resizebox{0.4\textwidth}{!}{\includegraphics{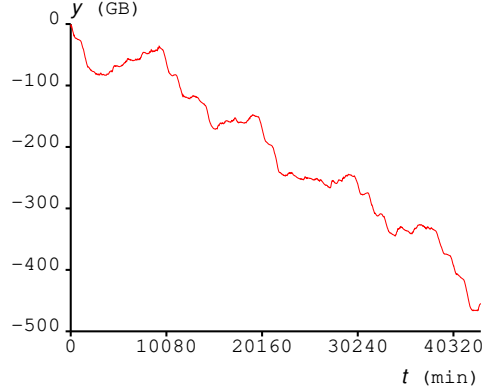}}
\caption{A modified profile of the packet flow corresponding to 
Fig.~\ref{profile.eps}.  The horizontal axis denotes time (minute) and
the vertical one does accumulated packet fluctuation around the
average (Giga Byte).  Weekly periodicity (10080 min) still remains.}
\label{profile-mod.eps}
\end{center}
\end{figure}

\begin{figure}[tb]
\begin{center}
\resizebox{0.4\textwidth}{!}{\includegraphics{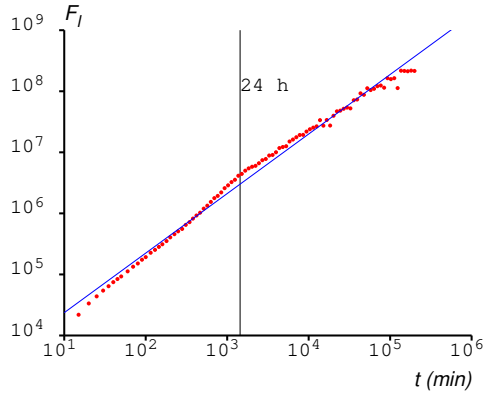}}
\caption{Dependence of the standard deviation $F(l)$ of the modified data
on segment length $l$ for MRTG data.  The exponent is $\alpha\sim 0.96$.}
\label{dfa-mod.eps}
\end{center}
\end{figure}

We will confirm, in this section, that the observed Internet traffic
consists of daily periodicity with power-law fluctuations.  We apply
the same procedure applied in our previous analysis of expressway
vehicle traffic\cite{Tadaki2006}.

First we define the daily trend $\tilde{q}_\mathrm{daily}(\tau)$ using
the same method for analyzing vehicle traffic as in the previous work
on vehicle traffic.  The daily trend is defined by
\begin{equation}
\tilde{q}_\mathrm{daily}(\tau)
=\frac{1}{D}\sum_{d=0}^{D-1}q(d\times T_\mathrm{day}+\tau),
\label{daily-trend}
\end{equation}
where $0<\tau<T_\mathrm{day}=24\times60\ \mathrm{min}$ and $D$ is the
number of days in the data.  
%The raw data and the corresponding daily
%trend are shown in Fig.~\ref{daily-trend.eps}.  
Then the packet flow $q(t)$ is replaced with the modified packet flow
$q'(t)$
\begin{equation}
q'(t)=q(t)-\tilde{q}_\mathrm{daily}(t \bmod T_\mathrm{day}).
\label{dailyExtract}
\end{equation}

The profile $y'(t)$ for the modified packet flow $q'(t)$ is defined in
the same way as in eq.~(\ref{ProfileDefine})
\begin{equation}
y'(t)=\sum_{i=0}^{t}\left[q'(i)-\left\langle q'\right\rangle\right].
\end{equation}
The profile $y'(t)$ for the modified flow is shown in
Fig.~\ref{profile-mod.eps}.  It does not show daily periodicity no
longer.  We still recognize weekly periodic behavior in the modified
profile $y'(t)$.

The standard deviation eq.~(\ref{Deviation:1}) for the modified flow
is evaluated (Fig.~\ref{dfa-mod.eps}).  The bending point at a day
long almost disappears. The power-law fluctuation property is clearly
shown.  Namely Internet packet flow is a daily periodic one with
power-law fluctuations.  This property is not restricted in smaller
time scale than a day as observed previous works.  It extends to
longer time scale than a month.

The exponent of the fluctuation shown in Fig.~\ref{dfa-mod.eps} is
$\alpha\sim 0.96$.  It means that the power spectrum behaves $P(k)\sim
k^{-\beta}$ with $\beta\sim0.92$.  Namely, Internet traffic contains
$1/f$ fluctuations.

\begin{figure}[tb]
\begin{center}
\resizebox{0.4\textwidth}{!}{\includegraphics{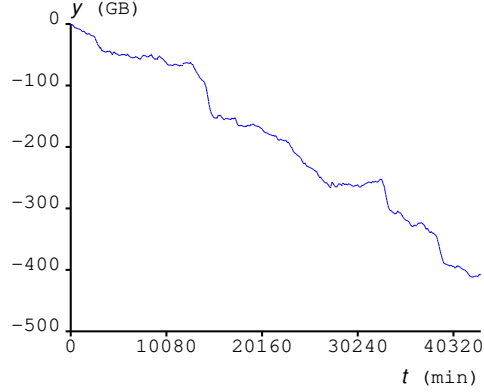}}
\caption{A profile of the packet flow without daily and weekly 
average periodicity.  The horizontal axis denotes time (minute) and
the vertical one does accumulated packet fluctuation around the
average (Giga Byte).  Any periodic behavior is not recognized.}
\label{profile-mod-weekly.eps}
\end{center}
\end{figure}

\begin{figure}[tb]
\begin{center}
\resizebox{0.4\textwidth}{!}{\includegraphics{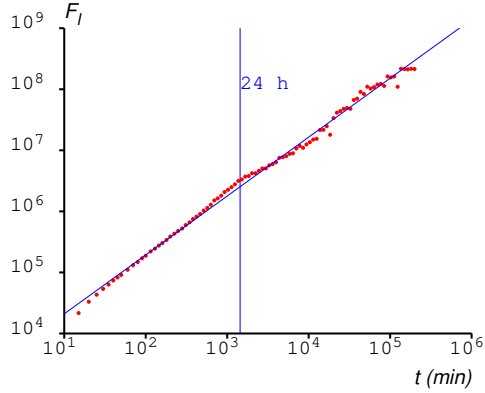}}
\caption{Dependence of the standard deviation $F(l)$ of the modified
data without daily and weekly average periodicity on segment length
$l$ for MRTG data.  Weekly periodicity does not affect the standard
deviation $F(l)$.  The exponent is $\alpha\sim 0.96$.}
\label{dfa-mod-weekly.eps}
\end{center}
\end{figure}

The modified packet flow $q'(t)$, which is defined by extracting daily
average periodicity from the original packet flow $q(t)$, still has
weekly periodicity as shown in Fig.~\ref{profile-mod.eps}.  We finally
discuss the effects of weekly periodicity remaining in $q'(t)$ on the 
standard deviation shown in Fig.~\ref{dfa-mod.eps}.

We can defined a new modified packet flow $q''(t)$ by extracting the
remaining weekly periodicity in $q'(t)$ using the same procedure
employed in eq.~(\ref{dailyExtract}).  The profile of the sequence,
whose daily and weekly average periodicity is extracted, is shown
in Fig.~\ref{profile-mod-weekly.eps}.  The profile seems not to
contain any periodicity.  The standard deviation corresponding to
the profile in Fig.~\ref{profile-mod-weekly.eps} is shown in
Fig.~\ref{dfa-mod-weekly.eps}.  There are no new features in
comparison with Fig.~\ref{dfa-mod.eps}.  The exponent $\alpha\sim0.96$
equals to that in Fig.~\ref{dfa-mod.eps}.  Namely the weekly
periodicity is not important to discuss the long-range correlation.
This supports that the packet flow $q(t)$ is a daily periodic one
with power-law fluctuations.

\section{Summary and Discussion}\label{Summary}
\begin{figure}[tb]
\begin{center}
\resizebox{0.4\textwidth}{!}{\includegraphics{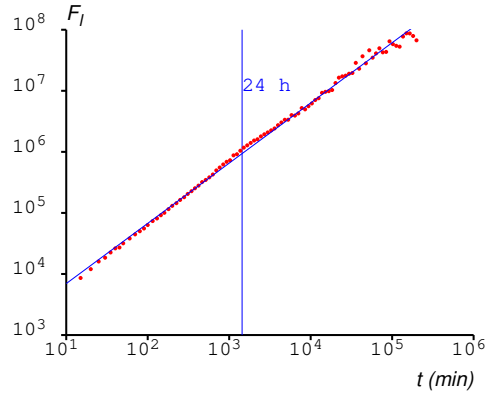}}
\caption{Dependence of the standard deviation $F(l)$ of the modified data
on segment length $l$ for outgoing MRTG data.  The exponent is $\alpha\sim 0.99$.}
\label{dfa-mod-out.eps}
\end{center}
\end{figure}

The Internet has been attracting scientific interests because of its
complex and autonomous properties.  The assumption that Internet
traffic obeys a Poisson distribution, has lost its validity.
Researchers have been revealing scale-free properties of the Internet
in various features.  This study analyzes the time sequence of Internet
packet flow observed using MRTG and discuss the power-law properties.

We employ the detrended fluctuation analysis (DFA) to analyze the
long-range correlation in the observed data of Internet flow.  The
result shows the typical feature of periodic behavior with power-law
fluctuations.  It also shows the dominant periodicity is one-day long.

To confirm that Internet packet flow is a mixuture of daily periodic
motion and power-law fluctuations, we apply the same method employed
for analyzing the expressway vehicle traffic in our previous work.
The modified packet flow data is defined by extracting the average
daily periodicity from the raw data.  By DFA applied on the modified data,
the power-law fluctuation is clearly observed in longer time scale
than a month.

We analyzed the incoming packet flow into Saga University.  The amount
of the outgoing packets is two or three times smaller than that of the
incoming flow.  The property of the power-law fluctuation is the same
as that of incoming flow.  By DFA applied on the modified outgoing
packet flow, the standard deviation behaves in the same way as for the
incoming packet flow (Fig.~\ref{dfa-mod-out.eps}).  The power-law
fluctuation seems to be general in Internet flow.

%Exclusion and queuing mechanisms discussed in \S~\ref{Introduction}
%will affect power-law properties in small time-scales.  The Internet
%contains various exclusion and queuing mechanisms with longer
%time-scales.  A sendmail system, for example, has such mechanisms.  If
%a message is not delivered to the destination, a sendmail system tries
%to re-send it after 15 or 30 minutes.  Undelivered messages will be
%kept for one week.  These queuing mechanisms with long time-scales
%will be effective to generate power-law fluctuations with long
%time-scales.

Exclusion and queuing mechanisms discussed in \S~\ref{Introduction}
will affect power-law properties in small time-scales.  Random
fluctuations in shorter time-scales than a second pile up by exclusion
and queuing effects in network systems.  And those effects generate
power-law fluctuations in times-scales of hours.  The Internet also
contains various exclusion and queuing mechanisms with longer
time-scales than a day.  A sendmail system, for example, has such
mechanisms.  First a user sends his e-mail messages to the sendmail
server of his organization.  The server tries to deliver the message
to the server of message's destination.  If the message is not
delivered to the destination, the server tries to re-send it after 15
or 30 minutes.  The server will keep undelivered messages for one week
and repeat to try to send them.  If a queued message finally fails to
be sent after one week, the server generates a message of an
undelivered event for the sender of the message.  These queuing
mechanisms with long time-scales will be effective to generate
power-law fluctuations with long time-scales.  And the retrying
behavior in the sendmail system also affects the performance of
computer and network systems and generates long-range correlation.

%The power-law fluctuations longer than a month, however, seems to be
%independent of network mechanisms.  Exclusion and queuing mechanisms
%in human activities may be important for power-law properties of the
%Internet.

The power-law fluctuations longer than a month, however, seems
difficult to be discussed depending only on network mechanisms.
Exclusion and queuing mechanisms in human activities may be important
for power-law properties of the Internet.  In human activities, we
also have finite queue to handle jobs.  We postpone some jobs
depending on their importance.  The delay of one's job causes
avalanches of delays in subsequent jobs scheduled by others.  The
Internet is one of the main communication tools of modern working
environment.  So finite queues in human activities affect the
long-range correlation in the Internet.  Anyway we need some simple
models of network services, human activities, and their interaction.

The scale-free properties in the global structure of the Internet will
be another origin of the power-law fluctuations in Internet traffic.
We observe the outgoing packet flow also shows power-law fluctuations.
The structure of our university LAN is not scale free. It is a simple
hierarchy with a few layers.  So we need to study how the global
scale-free structure of the Internet affects the power-law properties
of packet flow at a peripheral gateway.

\section*{Acknowledgments}
A part of this work is financially supported by a Grant-in-Aid
No.~18500215 from the Ministry of Education, Culture, Sports, Science
and Technology, Japan.

\end{document}